\documentclass[aps,prb,preprint,groupedaddress,showpacs,floatfix]{revtex4}
\usepackage{graphicx}
\usepackage{color}

\begin{document}

\title{Control of spin injection by direct current in lateral spin valves}

\author{F\`elix Casanova}
  \email{casanova@physics.ucsd.edu}
  
\author{Amos Sharoni}
  \email{asharoni@physics.ucsd.edu}
  \altaffiliation{both authors, F.C. and A.S., have contributed equally to this work.}
\author{Mikhail Erekhinsky}
\author{Ivan K. Schuller}

\affiliation{Physics Department, University of California - San Diego, La Jolla, California 92093}
  
\date{\today}

\begin{abstract}
The spin injection and accumulation in metallic lateral spin valves with transparent interfaces is studied using d.c. injection current. Unlike a.c.-based techniques, this allows investigating the effects of the direction and magnitude of the injected current. We find that the spin accumulation is reversed by changing the direction of the injected current, whereas its magnitude does not change. The injection mechanism for both current directions is thus perfectly symmetric, leading to the same spin injection efficiency for both spin types. This result is accounted for by a spin-dependent diffusion model. Joule heating increases considerably the local temperature in the spin valves when high current densities are injected ($\sim$80--105 K for 1--2$\times10^{7}$A cm$^{-2}$), strongly affecting the spin accumulation.

\end{abstract}

%\keywords{}

\maketitle

\section{Introduction}
The generation and control of spin currents is a key ingredient in spintronics, which has as a goal the use of both the spin and charge degrees of freedom of the electron \cite{Zutic:2004}. As an example, ferromagnetic (FM) / non-magnetic (NM) lateral spin valves are powerful devices that decouple a pure spin current from an electrical current by taking advantage of a non-local geometry \cite{Johnson:1985, Jedema:2001, Jedema:2002}. With such hybrid nanostructures, a spin-polarized current has been injected into a metal \cite{Johnson:1985, Jedema:2001, Jedema:2002, Jedema:2003, Valenzuela:2004, Kimura:2005, Garzon:2005, KimuraPRL:2007, Ji:2007}, a semiconductor \cite{Lou:2007} or a superconductor \cite{Beckmann:2004, Poli:2008}, leading to the observation of new fundamental phenomena (such as the spin Hall effect in metals \cite{Valenzuela:2006} or the crossed Andreev reflection in superconductors \cite{Beckmann:2004}) or to new possible applications (such as integrated spintronic circuits in semiconductors \cite{Dery:2007}). A physical understanding of the creation and manipulation of a spin current from an electrical spin injection is essential for the development of these spintronic devices, in which many device characteristics, such as geometry or materials properties \cite{Takahashi:2003, Johnson:2007} play a major role. The FM/NM interface conductivity is an important controlling parameter. The spin polarization of a current injected through a tunnel junction strongly decreases with applied bias \cite{Valenzuela:2004, Valenzuela:2005} limiting the maximum spin current density. However, a similar study for transparent interfaces is lacking. 

Most measurements of these non-local spin valves (NLSV) use an alternating current (a.c.) lock-in technique to extract the relatively small spin signal from the background noise. When using a.c., it is difficult to study the effects of the magnitude and direction of the injected charge current on the spin current. In addition, any information regarding offsets in the measurements are lost in the method. Therefore, achieving full control of the electrical spin injection and a better understanding of the physical phenomenon is not possible with a.c. methods. In this work, we present the use of direct current (d.c.) in NLSV measurements to control the generated spin current and to further understand the injection mechanism in transparent contacts as well as the origin of background signals appearing in experimental devices. We investigated the effect of both the current magnitude and current direction on spin injection and accumulation in metallic lateral spin valves with transparent contacts. We find experimentally that the spin accumulation in the NM reverses with reversal of the injected current (up to current densities $\geq 3\!\times\!10^{6}$A cm$^{-2}$), while keeping the same magnitude, indicating a symmetric injection mechanism. This result, in agreement with a spin-dependent diffusion model \cite{Valet:1993}, enables an electrical, magnetic-field-free control of spin currents. We identify the origin of two different background contributions to the NLSV measurements: inhomogeneous current distribution and thermoelectric effect due to Joule heating in the spin valve device. This Joule heating also increases the local temperature of the device, modifying the magnitude of the spin accumulation at high current densities.
\begin{figure}[tbp]
\begin{center}
\includegraphics[width=8.5 cm]{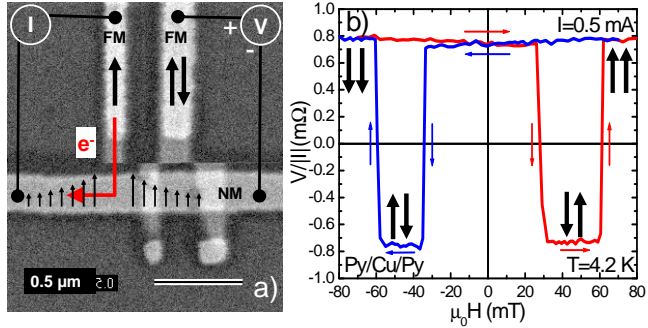}
\caption{\label{fig1} (Color online). a) Scanning electron microscope image of a lateral spin valve with a schematic illustration of spin injection, accumulation and detection in a non-local measurement. Thick vertical arrows indicate the state of the magnetization in the ferromagnetic electrodes. Thin vertical arrows represent the accumulation of injected spins in the non-magnetic strip. b) Normalized NLSV signal measured in a Py/Cu/Py lateral spin valve at 4.2 K and 0.5 mA with a "d.c. reversal" technique, while sweeping the magnetic field in the direction given by the thin arrows. Thick vertical arrows indicate again the magnetic alignment of the electrodes.}
\end{center}
\end{figure}

\section{Experimental details}
Lateral spin valve devices are fabricated using a two-angle shadow evaporation technique, which allows fabrication of the full device in-situ, necessary to obtain clean transparent interfaces. First, a suspended mask is created by e-beam lithography on a bilayer resist (500-nm-thick MMA / 200-nm-thick PMMA) on top of a Si substrate. Two FM electrodes are then deposited by e-beam evaporation of 35 nm of Py or Co at a base pressure of $\sim 3\times10^{-10}$ mbar at a $75^{\circ}$ angle from the normal to the substrate. Next, without breaking vacuum, 120 nm of Cu or Al are deposited at the normal to the substrate to form a NM strip. Here, we prepared several Py/Cu/Py and Co/Al/Co lateral spin valves. Figure \ref{fig1} (a) shows a scanning electron microscopy (SEM) image of one device. The width of the NM strip is 230 nm. The resistivities for Cu and Al at 4.2 K are 1.67 $\mu\Omega$ cm and 4.83 $\mu\Omega$ cm, respectively. Different widths of the FM electrodes, typically 90 and 160 nm, ensure different switching fields. The edge-to-edge distance between them was varied from 200 nm to 1500 nm. The interface resistance multiplied by the interface area at 4.2 K is typically $3.4\times10^{-4}\Omega$ $\mu$m$^{2}$ for Py/Cu and $1.0\times10^{-2}\Omega$ $\mu$m$^{2}$ for Co/Al. Since all devices behave similarly, we present the results for a particular Py/Cu/Py and Co/Al/Co spin valve, unless otherwise stated. Electrical measurements in our devices were done using a d.c. current source and a nanovoltmeter. We have measured voltages for positive and negative current separately, subtracting them from the voltage measured at zero current to eliminate any instrumental offset. With the same setup, we have also measured voltages using a "d.c. reversal" method, equivalent to an a.c. lock-in technique \cite{DeltaWP:note}. This allows us to compare a.c. with d.c. measurements. In the "d.c. reversal" measurement (as in any a.c. measurement), although it gives a better signal-to-noise ratio, any information about the effect of the direction of the current is lost.

In all our measurements we use the non-local geometry, in which spin-polarized electrons are injected from a FM electrode (injector) into the NM strip, where it produces a non-equilibrium spin accumulation [Fig. \ref{fig1} (a)], i.e., there is a splitting of the electrochemical potential for spin-up and spin-down electrons. The electrical current flows only in the left side, whereas the spin accumulation diffuses in both directions along the NM strip (a pure spin current). This diffusion occurs along a characteristic length scale - the spin diffusion length ($\lambda$)\cite{Bass:2007}. A second FM electrode (detector) placed on the right side detects the spin accumulation, i.e., the difference in electrochemical potentials between spin-up and spin-down sub-band gives rise to a voltage difference between the detector and the NM strip \cite{Johnson:1985, Jedema:2001, Takahashi:2003}. This voltage shows bipolar switching when the relative magnetic alignment of the FM electrodes changes from parallel (P) to antiparallel (AP), a pure spin valve effect. The experimentally measured voltage may contain background contributions (shifts from ideal bipolar behavior), as is often observed in transparent \cite{Ji:2007} and tunneling \cite{Jedema:2002, Valenzuela:2004} spin valves using a.c. measurements. The latter voltage, normalized to the magnitude of the current ($V/\left|I\right|$), is the NLSV signal. The NLSV signal difference, $(V_P-V_{AP})/\left|I\right|=\Delta V/\left|I\right|$, which does not depend on the background but only on the pure spin valve effect, is proportional to the spin accumulation under the detector \cite{Garzon:2005}. 

From the $\Delta V/\left|I\right|$ values at different edge-to-edge distances between electrodes and using Eq. \ref{eq1} (see below), with values for the spin diffusion length of the FM $\lambda_{Py}$=5 nm \cite{Dubois:1999, Jedema:2001, KimuraPRL:2008, Bass:2007} and $\lambda_{Co}$=36 nm \cite{Dubois:1999, Jedema:2003, Bass:2007}, we can derive the spin polarization of the FM ($\alpha_F$) and the spin diffusion length of the NM ($\lambda_N$) \cite{Jedema:2001, Jedema:2003, Kimura:2005, KimuraPRL:2007}. For the Py/Cu/Py spin valves, we obtain $\alpha_{Py}$=0.29 and $\lambda_{Cu}$=395 nm and for the Co/Al/Co spin valves, $\alpha_{Co}$=0.08 and $\lambda_{Al}$=425 nm, at 4.2 K. The spin diffusion lengths are in agreement with values previously reported \cite{Bass:2007}. The spin polarization of Py is also in good agreement with literature \cite{Jedema:2001, Jedema:2003, Kimura:2005, Dubois:1999,  KimuraPRL:2008}, whereas the value for Co is much smaller than $\sim$0.5 usually obtained \cite{Lee:1995, Tsymbal:1996, Doudin:1996, Piraux:1998}. This large reduction of $\alpha_{Co}$ has also been observed in previous experiments using lateral spin valves \cite{Jedema:2003}. Note that $\alpha_{F}$ is the intrisic spin polarization of the FM only in an ideal case and any interface or geometry effect in the spin injection will be reflected in the fitted value of $\alpha_{F}$. A discussion of the origin of the discrepancy in $\alpha_{Co}$ is given in Ref. \onlinecite{Jedema:2003}.

\begin{figure}[tbp]
\begin{center}
\includegraphics[width=8.5 cm]{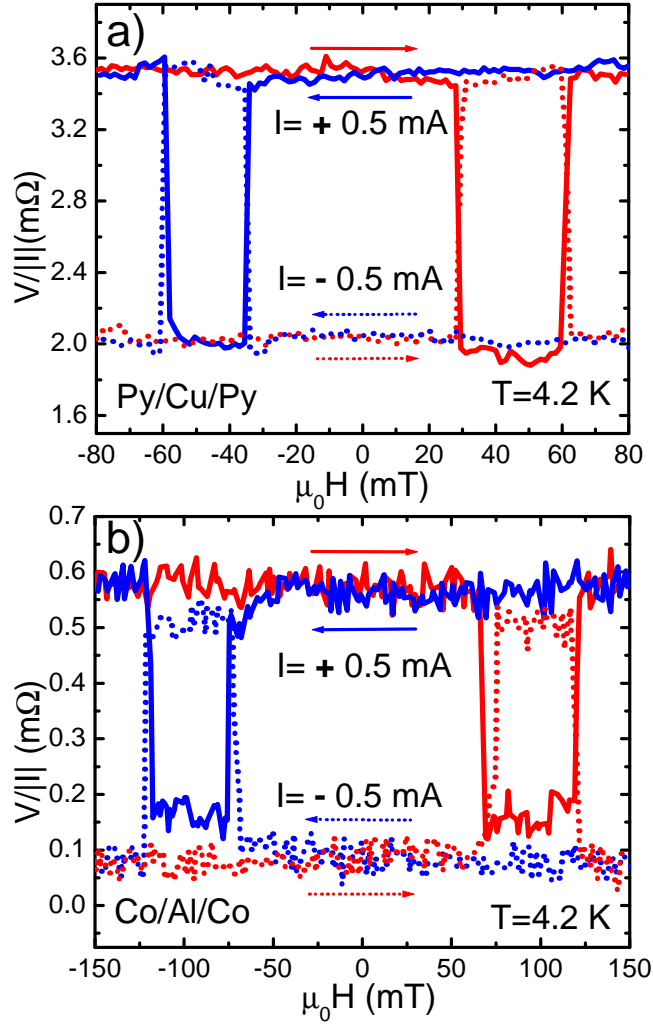}
\caption{\label{fig2} (Color online). Normalized NLSV signal measured in a a) Py/Cu/Py and b) Co/Al/Co lateral spin valve at 4.2 K and 0.5 mA while sweeping the magnetic field in the direction given by the horizontal arrows. Signal measured at positive and negative d.c. current is plotted as a solid and a dotted line, respectively.}
\end{center}
\end{figure}

\section{results and discussion}
\subsection{Effect of current direction: results}
\label{results}
Figure \ref{fig1}(b) shows an example of a NLSV measurement, using the "d.c. reversal" method for the Py/Cu/Py lateral spin valve, as a function of the applied magnetic field parallel to the FM electrodes. A clear bipolar switching of the NLSV signal is observed when the magnetization of FM electrodes changes from P to AP, resulting in a difference of $\Delta V/\left|I\right|$=1.5 m$\Omega$. We note that, in some of our spin valves (with nominally the same geometry), the NLSV signal measured with the "d.c. reversal" method has an offset that shifts the bipolar behavior up. The origin of this offset has been identified and is discussed in sec. \ref{background}.

To study the effect of the injected current direction, we repeated the previous NLSV measurement, using  positive (+I) and negative (-I) currents separately. Figure \ref{fig2} (a) shows the result for the Py/Cu/Py device. The NLSV signal is completely reversed for opposite currents, i.e., the NLSV signal for P (AP) alignment with +I is the same as the one for AP (P) alignment with -I. The magnitude of $\Delta V/\left|I\right|$ (1.5 m$\Omega$) is the same as in the "d.c. reversal" result, although a large constant offset ($\sim$2.8m$\Omega$) is observed and denoted as a "background". In the devices in which an offset in the NLSV signal was observed with the "d.c. reversal" method, a different background is observed for NLSV signals obtained with +I and with -I (see for example Fig. \ref{fig2} (b) for Co/Al/Co). The difference between the backgrounds with +I and -I is proportional to the offset observed with the "d.c. reversal" method. Nevertheless, the magnitude of $\Delta V/\left|I\right|$ remains constant for both current directions.

\begin{figure}[tbp]
\begin{center}
\includegraphics[width=8.5 cm]{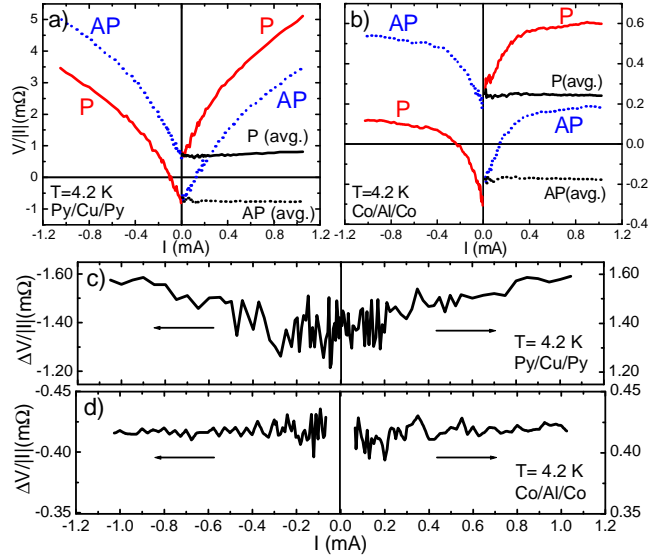}
\caption{\label{fig3} (Color online). Normalized NLSV signal measured in a a) Py/Cu/Py and b) Co/Al/Co lateral spin valve at 4.2 K as a function of d.c. current for a parallel (P, red solid lines) and antiparallel (AP, blue dotted lines) magnetic alignment of the electrodes. The same NLSV signal averaged for positive and negative d.c. current, $\frac{V(+I)-V(-I)}{2\left|I\right|}$, is plotted as a solid (dotted) black line for a parallel (antiparallel) configuration as a function of the absolute value of the current. c) and d) shows NLSV signal difference, $(V_P-V_{AP})/\left|I\right|=\Delta V/\left|I\right|$, as a function of d.c. current, calculated from data in a) and b), respectively. Note the opposite sign in the vertical scale for positive and negative d.c. currents.}
\end{center}
\end{figure}

Figures \ref{fig3} (a) and (b) show the NLSV signal for the P and AP magnetic alignment of the electrodes separately, as a function of d.c. current from -1 to +1 mA ($\sim 3\!\times\!10^{6}$A cm$^{-2}$). In the Py/Cu/Py device, for any current magnitude, the NLSV signal is the same when both the magnetic configuration and direction of current are reversed, i.e., the signal measured for P alignment with +I (-I) is identical to the one for AP alignment with -I (+I). Therefore, the NLSV signal difference $\Delta V/\left|I\right|$ has the same magnitude and opposite sign for positive and negative currents [plotted in Fig. \ref{fig3} (c)]. The origin of the observed increase of the NLSV signal for both P and AP alignments with the current magnitude is discussed in sec. \ref{background}. The NLSV signal for the Co/Al/Co device [Fig. \ref{fig3} (b)] behaves similarly to the previous device, except for an additional constant offset that shifts the signal up for +I and down for -I, also observed in Fig. \ref{fig2} (b) and discussed in sec. \ref{background}. However, also here, $\Delta V/\left|I\right|$ has the same magnitude and opposite sign for +I and -I, as shown in Fig. \ref{fig3} (d).

\subsection{Effect of current direction: discussion}
\begin{figure}[tbp]
\begin{center}
\includegraphics[width=8.5 cm]{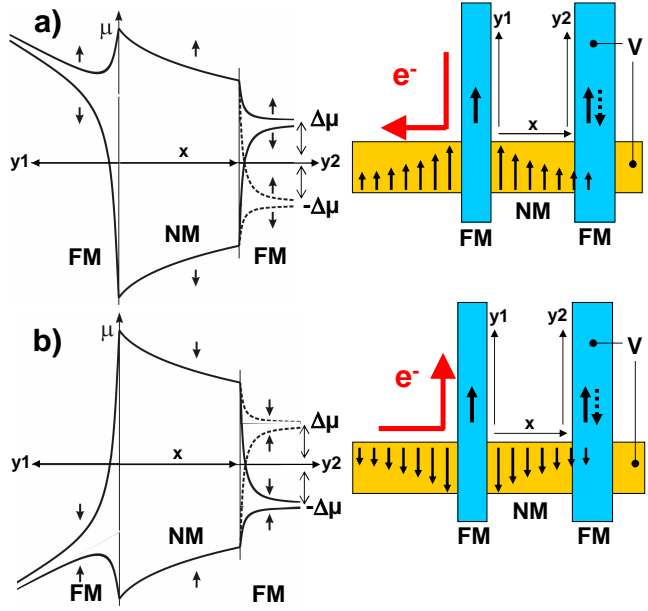}
\caption{\label{fig4} (Color online). The spatial dependence of the electrochemical potential $\mu$ for spin-up and spin-down electrons in the non-local configuration of a lateral spin valve [FM injector (y1, left), normal metal (x, middle) and FM detector (y2, right)] when injecting a) negative and b) positive d.c. current. Solid (dashed) lines in the FM detector correspond to $\mu$ when the magnetic alignment is parallel (antiparallel). The corresponding schematic illustration of spin injection, accumulation and detection for negative and positive d.c. current is drawn to the right.}
\end{center}
\end{figure}

In order to understand this symmetric behavior, one has to consider the detailed mechanism of spin injection and accumulation. The basic model for spin transport in the diffusive regime formulated by \citet{Valet:1993} is generally used in lateral spin valves \cite{Jedema:2001, Jedema:2003, Takahashi:2003, Kimura:2005, KimuraPRL:2007} since the mean free path of electrons is shorter than the usual device dimensions. By solving the one-dimensional spin-dependent diffusion equation given by this model \cite{Valet:1993, Jedema:2003, Takahashi:2003, Kimura:2005} for the non-local geometry of our device, we obtain the electrochemical potential for the spin-up and spin-down electrons as a function of the position (Fig. \ref{fig4}). The NLSV voltage built at the NM-FM detector interface is given by:
\begin{equation} 
V=I\frac{\alpha_{F_1}\alpha_{F_2}R_{N}}{\left(2+\frac{R_{N}}{R_{F}}\right)^{2}e^{d/\lambda_{N}}-\left(\frac{R_{N}}{R_{F}}\right)^{2}e^{-d/\lambda_{N}}} 
\label{eq1}
\end{equation}
where $R_{N}=2\lambda_{N}\rho_{N}/S_{N}$ and $R_{F}=2\lambda_{F}\rho_{F}/S_{F}(1-\alpha_{F}^2)$ are the spin resistances (a measure of the difficulty for spin mixing \cite{Kimura:2005}) for the NM and FM, respectively, $\lambda_{N,F}$ are the spin diffusion lengths, $\rho_{N,F}$ are the resistivities, $S_{N,F}$ are the cross-sectional areas, $d$ is the distance between FM electrodes and $\alpha_{F_1}$, $\alpha_{F_2}$ are the spin polarizations of each FM electrode, which have the same magnitude ($\alpha_{F}$) and the same sign for a P alignment, but opposite signs for an AP alignment.

The result is different if a negative [see Fig. \ref{fig4} (a)] or a positive [Fig. \ref{fig4} (b)] charge current is injected. When electrons are injected from the FM to the NM (negative currents), most of them are majority-spin electrons \cite{Majority:note}, causing a spin accumulation of majority spins in the NM. A parallel alignment of the FM detector with the injector will cause a negative voltage [$V=\Delta\mu /(-e)$] at the interface between the NM and the FM detector, because the majority spins at the detector have the same orientation as the ones accumulated in the NM. In contrast, an antiparallel alignment will produce a positive voltage [$V=-\Delta\mu /(-e)$] because the majority spins at the detector have the opposite orientation as the ones accumulated in the NM [see Fig. \ref{fig4} (a)]. When electrons are injected from the NM to the FM (positive currents), mostly majority-spin electrons are injected into the FM electrode, leaving a spin accumulation of minority spins in the NM. In this case, a parallel alignment will cause a positive voltage, whereas an antiparallel alignment will produce a negative voltage [see Fig. \ref{fig4} (b)]. It is worth noting that Eq. \ref{eq1} captures the sign change of the pure spin valve effect with current and magnetic alignment as we observe experimentally.

Since all parameters of the lateral spin valve (geometry and materials) remain the same when the current direction is reversed, any variation in $\Delta V/\left|I\right|$ would be due to a change in the spin injection efficiency, i.e., the spin polarization of the injected current at the interface. Accordingly, the spin injection efficiency is the same when injecting current from a FM to a NM and from a NM to a FM, causing exactly the opposite spin accumulation in the NM. This demonstrates that the injection mechanism in transparent junctions is perfectly symmetric for opposite spin types up to 1 mA ($\sim 3\!\times\!10^{6}$A cm$^{-2}$). 

This result, although predicted by Eq. \ref{eq1}, has been overlooked due to the use of a.c. currents and has not been proven experimentally before. Even though it may look straightforward, this symmetry of spin injection in metals using transparent contacts is in contrast with injection across tunnel junctions into metals \cite{Valenzuela:2004, Valenzuela:2005} or semiconductors \cite{Lou:2007}, where a spin polarization asymmetry is observed for opposite biases. 

More importantly, these results show that we are able to manipulate the sign of the spin accumulation (i.e., the spin-type of the pure spin current) while keeping the same magnitude by only changing the direction of the electrical current, without changing the magnetic configuration of the spin valve (i.e., without application of an external magnetic field).

\subsection{Origin of backgrounds}
\label{background}
The origin of the backgrounds present in the NLSV signal can be inferred from their current dependences. In all devices, as the one measured in Fig. \ref{fig3} (a), the effects of the current on the background for any magnetic alignment of the electrodes (P or AP) are as follows: the background disappears when the current tends to zero (and the ideal bipolar switching, arising from the pure spin valve effect, occurs), increases with the absolute value of the d.c. current and is independent of current direction. To confirm these facts, we simulate an a.c. measurement by averaging the positive and negative current branches of the NLSV signal as $\frac{V(+I)-V(-I)}{2\left|I\right|}$. In this averaged NLSV signal [also plotted in Fig. \ref{fig3} (a) for P and AP alignments as a function of the absolute value of the d.c. current] the background is completely eliminated, which demonstrates that it cannot be observed in conventional a.c. measurements ("d.c. reversal" or lock-in). The fact that the background is independent of the current direction indicates that the origin is a thermoelectric effect due to Joule heating. The heat generated in the injecting junction dissipates along the NM strip, producing a temperature gradient between the detecting junction, where the temperature is higher, and the ends of the detecting FM electrode and the right side of the NM strip, where the temperature remains at 4.2 K. Therefore, a thermoelectric voltage is generated at the detecting loop, similar to the voltage measured in a thermocouple. This process is very sensitive, in which the details of the specific device such as the Seebeck coefficients of the materials or the interface resistance of the junctions, the thermal coupling to the substrate or the exchange gas are relevant, resulting in different backgrounds for different devices [compare, for example, Figs. \ref{fig3} (a) and (b)]. Therefore, no systematic correlation of the current-dependence of the background with the type of spin valve (Py/Cu/Py or Co/Al/Co) or geometric parameters is observed. It is worth noting that the temperature-dependent background in the NLSV signal observed in Ref. \onlinecite{Garzon:2005}, which we also observe in our devices, cannot be removed by a.c. methods. Therefore it has a different origin than the current-dependent background we observe. We show that the current-dependent background is completely removed in an a.c. measurement [both with a direct measurement, see Fig. \ref{fig1} (b), or by averaging the results of positive and negative d.c. currents, as shown in Figs. \ref{fig3} (a) and \ref{fig3} (b)], indicating a Joule heating origin.

A different background in the NLSV signal is also present in some devices (independently of the type of spin valve), as the one shown in Fig. \ref{fig3} (b): it is constant with current, being positive for positive currents and negative for negative currents, therefore having an ohmic behavior. The averaged NLSV signals for P and AP alignments, plotted in Fig. \ref{fig3} (b), are shifted up because this background is not eliminated. This background is thus the offset we observe for some samples in the "d.c. reversal" technique (see sec. \ref{results}) and also the one observed in conventional a.c. measurements \cite{Ji:2007}. The observation that it has an ohmic character and varies substantially from device to device is in agreement with a recent report \cite{Johnson:2007}. This report has identified the background to originate from an inhomogeneous current distribution which depends on the detailed device geometry. 

\begin{figure}[tbp]
\begin{center}
\includegraphics[width=8.5 cm]{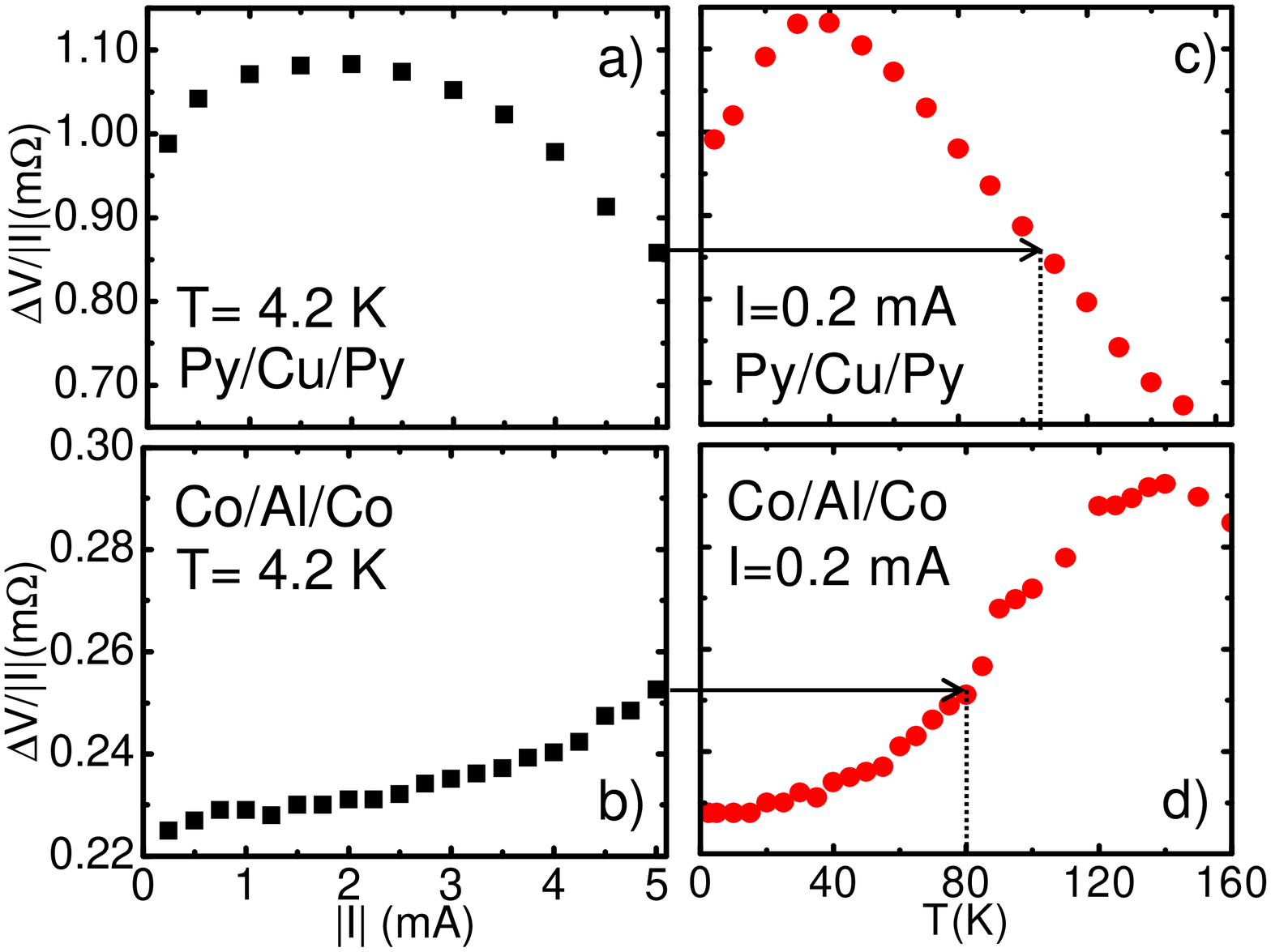}
\caption{\label{fig5} (Color online). NLSV signal difference, $(V_P-V_{AP})/\left|I\right|=\Delta V/\left|I\right|$, measured in Py/Cu/Py [a) and c)] and Co/Al/Co [b) and d)] lateral spin valves with a "d.c. reversal" technique as a function of the current magnitude at 4.2 K [a) and b)] and as a function of temperature at 0.2 mA [c) and d)]. Horizontal arrows and dotted vertical lines are used to estimate the local temperature of the lateral spin valves when 5 mA are applied. Note that the devices shown here are different from the ones shown in the previous figures.}
\end{center}
\end{figure}

\subsection{Effect of the current magnitude and temperature}
Finally, we show the dependence of the NLSV signal difference $\Delta V/\left|I\right|$ on the magnitude of the current up to 5 mA measured with a "d.c. reversal" technique in a Py/Cu/Py [Fig. \ref{fig5} (a)] and a Co/Al/Co [Fig. \ref{fig5} (b)] device. From the spin-dependent diffusion model (Eq. \ref{eq1}), one would expect a constant $\Delta V/\left|I\right|$ for any current, although this prediction fails in lateral spin valves with tunnel barriers\cite{Valenzuela:2004}, in which $\Delta V/\left|I\right|$ decays above 1--2$\times10^{4}$A cm$^{-2}$  because injected electrons tunnel into hot states with reduced polarization \cite{Valenzuela:2005}. Surprisingly, $\Delta V/\left|I\right|$ also varies with the injected current in transparent contacts: in Py/Cu/Py devices it first increases and then decreases with increasing the current magnitude [Fig. \ref{fig5} (a), showing a different device, nominally identical] and it only increases with increasing the current magnitude in Co/Al/Co devices [Fig. \ref{fig5} (b), showing a different device, nominally identical]. $\Delta V/\left|I\right|$ as a function of the temperature for the same Py/Cu/Py [Fig. \ref{fig5} (c)] and Co/Al/Co [Fig. \ref{fig5} (d)] devices is also measured, yielding the same dependence as for the injected current. Therefore, the effects observed in Fig. \ref{fig5} (a) and (b) arise from a temperature increase of the device with the injected current, giving yet another experimental evidence that Joule heating occurs in lateral spin valves. The same dependence of $\Delta V/\left|I\right|$ with the temperature has been recently reported for Py/Cu/Py devices, which is attributed to the effect of temperature on the spin diffusion length of Cu \cite{KimuraPRL:2008}. From Fig. \ref{fig5}, we can estimate an increase of the local temperature to $\sim$105 K in the Py/Cu/Py device and $\sim$80 K in the Co/Al/Co device, when 5 mA (1--2$\times10^{7}$A cm$^{-2}$) are injected at 4.2 K.

\section{Conclusions}

In conclusion, we have studied systematically the effect of the direction and magnitude of a d.c. current on the spin injection and accumulation in metallic lateral spin valves with transparent junctions. We find that up to a current density of $3\!\times\!10^{6}$A cm$^{-2}$ the injection mechanisms are perfectly symmetric when injecting electrons from a FM to a NM, accumulating majority spins, and from a NM to a FM, accumulating minority spins, which causes exactly the opposite spin accumulation. This pure electrical manipulation of the polarity of the spin current with d.c. current is relevant for future magnetic-field-free spintronic devices. These results can be explained by a spin-dependent diffusion model. The two backgrounds appearing in the NLSV measurements originate from an inhomogeneous current distribution (observed in usual a.c. techniques) and from a thermoelectric effect due to Joule heating (observed only in d.c. measurements). Since high current densities are preferred for practical spintronic effects \cite{KimuraPRL:2007, Ji:2007}, such as spin torque \cite{Yang:2008}, spin injection with transparent junctions offers a greater advantage than with tunnel junctions, although the temperature effects due to Joule heating must be taken into account in the spin accumulation magnitude.

\section*{ACKNOWLEDGMENTS}
The financial support of the U.S. D.O.E. is acknowledged. 

%\bibliography{spininjection}

\begin{thebibliography}{29}
\bibitem[{Maj()}]{Majority:note}
\bibinfo{note}{Majority- and minority-spin electrons are defined with spin
  parallel and antiparallel to the magnetization of the FM electrode,
  respectively. Therefore, when the magnetization of the FM electrode points up
  (down), the majority-spin electrons are spin-up (spin-down) and the
  minority-spin electrons are spin-down (spin-up).}
\end{thebibliography}

\end{document}